\documentclass[12pt]{article}
\usepackage{setspace}
\usepackage{booktabs}
\usepackage{listings}
\usepackage{amsthm,amsmath,amssymb,natbib}
\usepackage{tikz}
\usetikzlibrary{calc, matrix, arrows, shapes, positioning, fit, shapes.misc, shapes.geometric, calc, decorations.pathreplacing}
\usepackage{caption}
\usepackage{minibox}
\usepackage{subcaption}
\usepackage{enumitem}
\usepackage{adjustbox}
\usepackage[margin=3.2cm]{geometry}
\usepackage{algorithm, algpseudocode}
\title{Link prediction for  egocentrically sampled networks}
\author{Yun-Jhong Wu, Elizaveta Levina, and Ji Zhu}

\theoremstyle{definition}

\theoremstyle{remark}

\theoremstyle{plain}

\begin{document}
\maketitle
\begin{abstract}
Link prediction in networks is typically accomplished by estimating or ranking the probabilities of edges for all pairs of nodes.    In practice, especially for social networks, the data are often collected by egocentric sampling, which means selecting a subset of nodes and recording all of their edges.    This sampling mechanism requires different prediction tools than the typical assumption of links missing at random.  We propose a new computationally efficient link prediction algorithm for egocentrically sampled networks, which estimates the underlying probability matrix by estimating its row space.  For networks created by sampling rows, our method outperforms many popular link prediction and graphon estimation techniques. 
\end{abstract}

\section{Introduction}
Networks are a useful tool for representing connections or relations between individual units, and a large body of work spread over several disciplines including statistics has been devoted to their analysis.      In many real networks generally and particularly in social networks, the edges are recorded with noise and missing values.   These problems can be especially severe when the data are collected by survey, which is not uncommon in social studies.     Link prediction addresses this problem by denosing the observed network and/or predicting missing links.   Many techniques have been delveloped for this task;  see \cite{Liben-Nowell2007} and \cite{Lu2011} for reviews. 

An undirected network on $N$ nodes can be represented with a symmetric adjacency matrix $\mathbf{A}\in\{0,1\}^{N\times N}$ with $A_{ij}= A_{ji}= 1$ if there is an edge between nodes $i$ and $j$.  In statistical network analysis, we usually assume that the adjacency $\mathbf{A}$ is generated from an underlying probability matrix $\mathbf{P}$, with $A_{ij}$'s generated as independent Bernoulli random variables, with probability that nodes $i$ and $j$ are connected given by some unknown $p_{ij}$.  The assumption of independence is not always realistic in practice, but so far the vast majority of probabilistic models for networks rely on it, and it has been found to produce useful algorithms.  

The link prediction problem can be thought of as classifying pairs of nodes as ``linked'' and ``not linked'', frequently done on the basis of a score for each pair, with an estimate of $p_{ij}$'s providing a natural score function.  Thus link prediction naturally leads to the problem of estimating  $\mathbf{P}$ or, alternatively,  a monotone function of the probabilities if only a relative ranking of links is important.   This is closely related to the problem of  matrix completion. 

\subsection{Matrix completion}
Matrix completion techniques have the same goal of denosing and/or completing a data matrix, often through  low-rank approximation.  Formally, they solve the optimization problem 
\begin{align*}
\operatorname*{\mathrm{min}}_{\mathbf{X}}&\quad\ell(\Omega(\mathbf{X}),\Omega(\mathbf{P})) \\
\mathrm{subject~to} &\quad \mathrm{rank}(\mathbf{X})\leq r,
\end{align*}
where $\Omega$ is an entry-wise mask operator with $\Omega(X_{ij})=X_{ij}$ if the entry $p_{ij}$ is observed and $\Omega(X_{ij})=0$ otherwise, and $\ell$ is a loss function. In the context of link prediction, we estimate $\mathbf{P}$ based on the observed adjacency matrix $\mathbf{A}$, which corresponds to noisy rather than noiseless matrix completion (the latter requires that observed entries are preserved, and the binary $A_{ij}$ is not a good estimator of $p_{ij}$).    In this case, the problem is solved using the empirical loss function, 
\begin{align*}
\operatorname*{\mathrm{min}}_{\mathbf{X}}&\quad\ell(\Omega(\mathbf{X}),\Omega(\mathbf{A})) \\
\mathrm{subject~to} &\quad \mathrm{rank}(\mathbf{X})\leq r,
\end{align*}
where we assume $A_{ij}=p_{ij}+e_{ij}$ with $\mathbb{E}[e_{ij}]=0$ and $e_{ij}$'s are independent.  Many 
theoretical results has been obtained for both noiseless and noisy formulations, for example, \citep{Candes2010, Candes2010a,Keshavan2010,Davenport2013}, and especially relevant to us, the universal singular value thresholding approach was proposed by \cite{Chatterjee2012}.    However, all the above results 
require the assumption that links in the adjacency matrix are missing at random, with a constant missingness rate $1-\rho$.    In this case, entries of the observed adjacency matrix have the form  $A_{ij}^{obs}=M_{ij}A_{ij}$, where $M_{ij} \sim \mathrm{Bernoulli}(\rho)$ indicates whether the status of the pair $(i,j)$ is observed.  
 Note that in this formulation for link prediction specifically,  $M$ is not observed, and so it is not possible to distinguish a missing link from a true 0, whereas an observed 1 always represents a true link.   One can estimate $\mathbb{E}[A_{ij}^{obs}]=\rho\mathbf{P}$ and a score-based classification method is still valid for predicting links even if $\rho$ is unknown \citep{Zhao2013}.    

\subsection{Egocentric networks}
Our focus here is on predicting links in networks constructed by egocentric sampling.  Egocentric networks have been studied in the quantitative social sciences for several decades \citep{Freeman1982,  Almquist2012} and more recently in physics and computer science \citep[e.g.][]{Newman2003a, Mcauley2012}.   It has been pointed out that summary statistics of egocentric networks can be dramatically different from those of a randomly sampled population network, due to the different structure of the noise introduced by the sampling mechanism \citep{Marsden2002, Kogovsek2005}. 
 It is reasonable to expect that this different noise structure will also affect many of the existing link prediction algorithms.

%%%%%%%%%
Egocentric sampling is often carried out  through surveys that ask a sample of subjects to name the people they are connected to according to the definition used by the study.    We model this process as sampling $n$ people without replacement from a group of size $N$, and asking them to name all their connections, without any upper bound on the number.   This results in an \textit{egocentric sample} or \textit{ego-network} consisting of a random sample of $n$ rows from the full $N \times N$ adjacency matrix.  

\begin{figure}[h]
\begin{center}
\begin{subfigure}[b]{0.4\textwidth}
\centering
   \begin{tikzpicture}
      \node[circle,draw,inner sep=1pt] (1) at (2,1.6) {1};
      \node[circle,draw,inner sep=1pt] (2) at (3.5,2.5) {2};
      \node[circle,draw,inner sep=1pt] (3) at (3,5) {3};
      \node[circle,draw,inner sep=1pt] (4) at (2.2,3) {4};
      \node[circle,draw,inner sep=1pt] (5) at (5.1,4) {5};
      \node[circle,draw,inner sep=1pt] (6) at (1,3) {6};

      \draw[thick] (4) -- (6);
      \draw[thick] (4) -- (1) -- (2);
      \draw[thick] (3) -- (2) -- (4) -- (3) -- (5);
      
  \end{tikzpicture}
                \caption{Full network $G$\\ \hspace{\textwidth}\\\hspace{\textwidth}}
        \end{subfigure}
\begin{subfigure}[b]{0.4\textwidth}
\centering
   \begin{tikzpicture}
      \node[circle,draw,inner sep=1pt] (1) at (2,1.6) {1};
      \node[circle,draw,inner sep=1pt] (2) at (3.5,2.5) {2};
      \node[circle,draw,inner sep=1pt, ultra thick] (3) at (3,5) {3};
      \node[circle,draw,inner sep=1pt] (4) at (2.2,3) {4};
      \node[circle,draw,inner sep=1pt] (5) at (5.1,4) {5};
      \node[circle,draw,inner sep=1pt, ultra thick] (6) at (1,3) {6};
      \draw[dashed, gray, thick] (1) -- (4) -- (2) -- (1);
      \draw[thick] (4) -- (6);
      \draw[thick] (3) -- (2);
      \draw[thick] (4) -- (3) -- (5);
      
  \end{tikzpicture}
                \caption{Ego network $G_{ego}$ constructed by sampling nodes $\{3,6\}$; dashed edges are not observed.}
        \end{subfigure}
\begin{subfigure}[b]{0.4\textwidth}
   \begin{align*}
\mathbf{A}=\left(\begin{array}{cccccc}
0 & 1 & 0 & 1 & 0 & 0 \\
1 & 0 & 1 & 1 & 0 & 0 \\
0 & 1 & 0 & 1 & 1 & 0 \\
1 & 1 & 1 & 0 & 0 & 1 \\
0 & 0 & 1 & 0 & 0 & 0 \\
0 & 0 & 0 & 1 & 0 & 0 \\
\end{array}\right)
  \end{align*}
                \caption{Adjacency matrix of G}
        \end{subfigure}
\begin{subfigure}[b]{0.4\textwidth}
   \begin{align*}
\mathbf{A}_{ego}=\left(\begin{array}{cccccc}
? & ? & 0 & ? & ? & 0 \\
? & ? & 1 & ? & ? & 0 \\
0 & 1 & 0 & 1 & 1 & 0 \\
? & ? & 1 & ? & ? & 1 \\
? & ? & 1 & ? & ? & 0 \\
0 & 0 & 0 & 1 & 0 & 0 \\
\end{array}\right)
  \end{align*}  
                \caption{Observed adjacency matrix}
        \end{subfigure}
        \caption{A toy example of egocentric sampling.} 
\label{ego:fig:ego_net}
\end{center}
\end{figure}

Formally, suppose that the network $G=(V,E)$ has the node set $V=\{1,\dots, N\}$ and the edge set $E$ with $|E| = m$.  
We sample nodes $\mathcal{I}=\{i_1,\dots,i_n\}\subset V$, and observed the ego-network $G_{ego}=(V_{ego},E_{ego})$, where $V_{ego}=V$, and $E_{ego}=\cup\{(u,v)\in E:  u \in I\}$.   See Figure \ref{ego:fig:ego_net} for an illustration.   Equivalently, when the $i$ node is sampled, we observe the $i$-th row and column of $\mathbf{A}$.

Related work on low rank approximations includes the CUR decomposition \cite{Mahoney2009}.  The purpose of the CUR algorithm is to find a matrix $\mathbf{U}$ such that $\mathbf{P}\approx\mathbf{CUR}$, where $\mathbf{C}$ and $\mathbf{R}$ are exactly the columns and rows sampled from $\mathbf{P}$.  This approach greatly reduces the time and space complexity for compressing a matrix, which was its main motivation. To obtain $\mathbf{U}$, the CUR algorithm solves a least squares problem, letting 
\[\mathbf{U}=\operatorname*{\arg\min}_{\mathbf{X}}\|\Omega(\mathbf{A})-\Omega(\mathbf{C}^\top\mathbf{X}\mathbf{R})\|_F^2,\]
where $\Omega(A_{ij})=1(i \mbox{ or } j \mbox{ is selected})$, and $\| \cdot \|_F$ isthe matrix Frobenius norm.    Later, we will also use $\| \cdot \|_2$ to denote the matrix spectral norm.    The theoretical foundation for the CUR decomposition \citep{Drineas2006a, Drineas2008} assumes the matrix to be noiseless, which for us would mean observing the probability matrix $\mathbf{P}$ directly instead of the adjacency matrix $\mathbf{A}$.    For link prediction, the CUR decomposition suffers from overfitting since we can always achieve $\min_\mathbf{X}\|\Omega(\mathbf{A})-\Omega(\mathbf{C}^\top\mathbf{X}\mathbf{R})\|_F^2=0$ by choosing $\mathbf{U}$ to be the pseudo-inverse of the intersection of $\mathbf{C}$ and $\mathbf{R}$. Moreover, to achieve an accurate CUR  approximation, it is essential to use importance sampling to sample $\mathbf{C}$ and $\mathbf{R}$ based on a probability distribution that is computed from the entire data matrix, which is feasible for matrix compression but not so much for egocentrically sampled social networks.   While we assume the rows of the egocentric networks are sampled uniformly without replacement, which may still not always be realistic, it is a big step in the right direction compared to assuming that links are missing uniformly at random.  
 
The rest of this paper is organized as follows. In Section \ref{sec:estimation}, we propose a new computationally efficient method for link prediction for egocentrically sampled networks based on a low rank approximation. The key idea is subspace estimation, since the observed rows allow us to estimate the approximate row space of the probability matrix $\mathbf{P}$.    Numerical evaluation on both synthetic and real networks and comparisons to benchmark link prediction methods are presented in Section \ref{sec:numerical_study}.  Section \ref{sec:discussion} concludes with discussion and future work.

\section{The subspace estimation algorithm for link prediction}\label{sec:estimation}
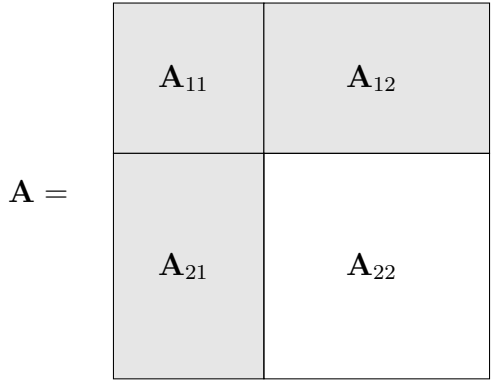
\begin{figure}
\begin{center}
\begin{tikzpicture}[
    all/.style={draw,align=center},
    insample/.style={all,fill=black!10},
    inrow/.style={all,minimum height=2cm},
    incol/.style={all,minimum width=2cm},
    outrow/.style={all,minimum height=3cm},
    outcol/.style={all,minimum width=3cm}
    ]
\begin{scope}
\node[] at (-3.5,0) {$\mathbf{A}=$};
\matrix (MAT) [matrix of nodes,nodes={all,anchor=center},column sep=-1\pgflinewidth,row sep=-1\pgflinewidth,inner sep=0pt]
{
   |[insample,inrow,incol]| $\mathbf{A}_{11}$ 
 & |[insample,outcol,inrow]| $\mathbf{A}_{12}$ \\
   |[insample,incol,outrow]| $\mathbf{A}_{21}$ 
 & |[outcol,outrow]| $\mathbf{A}_{22}$ \\   
  };
\end{scope}
\end{tikzpicture}
\end{center}
\caption{Grey:  observed blocks of the adjacency matrix.  White: the unobserved block. }\label{ego:fig:obs_adjmat}
\end{figure}

Without loss of generality, we can assume that the first $n$ nodes out of $N$ were selected, and the observed adjacency matrix can be partitioned into blocks $\mathbf{A}_{ij}$ for $i,j \in \{1,2\}$, where $\mathbf{A}_{11}\in\{0,1\}^{n\times n}$, $\mathbf{A}_{12}\in\{0,1\}^{n\times (N-n)}$, and $\mathbf{A}_{21}=\mathbf{A}_{21}^\top$, and the block $\mathbf{A}_{22}\in\{0,1\}^{(N-n)\times (N-n)}$ is not observed; see Figure \ref{ego:fig:obs_adjmat}.   The corresponding submatrices of $\mathbf{P}$ are denoted $\mathbf{P}_{ij}$ for $i,j=1,2$. We also define $\mathbf{A}_{in}=[\mathbf{A}_{11}~\mathbf{A}_{12}]_{n\times N}$ to be the sampled rows, with the corresponding probability sub-matrix $\mathbf{P}_{in}=[\mathbf{P}_{11}~\mathbf{P}_{12}]_{n\times N}$. 

\subsection{Estimation}

Our goal is to predict links between nodes that were not sampled, or equivalently to estimate $\mathbf{P}_{22}$.   We will do this by approximating the probability matrix $\mathbf{P}$ with a rank $r$ symmetric matrix $\mathbf{P}_r$. Suppose that we could directly sample rows from $\mathbf{P}$ instead of $\mathbf{A}$. The CUR decomposition uses the risk function
\[\ell(\mathbf{X};\mathbf{P})=\|\Omega(\mathbf{P})-\Omega(\mathbf{P}_{in}^\top\mathbf{X}\mathbf{P}_{in})\|_F , \]
with the corresponding estimator of $\mathbf{X}$ given by 
\[\mathbf{U}=\operatorname*{\arg\min}_{\mathbf{X}\in\mathbb{R}^{n\times n}}\|\Omega(\mathbf{A})-\Omega(\mathbf{A}_{in}^\top\mathbf{X}\mathbf{A}_{in})\|_F.\]
The solution to this optimization problem is $\mathbf{U}=\mathbf{A}_{11}^+$, where $\mathbf{A}_{11}^+$ is the Moore-Penrose pseudo-inverse of $\mathbf{A}_{11}$, 
 the $\mathbf{U}$ matrix in the standard CUR decomposition \citep{Mahoney2009}. However, its in-sample error is $\|\Omega(\mathbf{A})-\Omega(\mathbf{A}_{in}^\top\mathbf{A_{11}^+}\mathbf{A}_{in})\|_F=0$, and empirically we observed this estimator to often give poor predictions, suggesting it may suffer from overfitting.   Overfitting can be addressed through regularization, and a natural regularization to consider in this context is constraining the rank of $X$, computing instead 
\[\widetilde{\mathbf{X}}=\operatorname*{\arg\min}_{\mathrm{rank}(\mathbf{X})\leq r}\|\Omega(\mathbf{A}_{obs})-\Omega(\mathbf{A}_{in}^\top\mathbf{X}\mathbf{A}_{in})\|_F.\]
The resulting estimator of $\mathbf{P}$,  $\mathbf{A}_{in}^\top \tilde{\mathbf{X}}\mathbf{A}_{in}$ has rank $r$. We can think of it as an estimator of $\mathrm{row}(\mathbf{P}_r)$, which is the $r$-dimensional principal subspace of $\mathrm{row}(\mathbf{P})$.  
 However, solving this non-convex optimization problem is difficult.  Instead, we propose the following two-stage estimation procedure: we first estimate $\mathrm{row}(\mathbf{P}_r)$, which gives us an estimate of a principal subspace of $\mathrm{row}(\mathbf{P})$, and then, by considering \[\mathbf{P}_{11}^+=\operatorname{\arg\min}_{\mathbf{X}}\|\Omega(\mathbf{P})-\Omega(\mathbf{P}_{in}^\top\mathbf{X}\mathbf{P}_{in})\|_F,\] 
we construct a plug-in estimator of $\mathbf{P}$.
Specifically, we estimate $\mathbf{P}$ by the following steps:
\begin{enumerate}
\item Compute $\widetilde{\mathbf{P}}_{in}$, the best rank $r$ approximation to $\mathbf{A}_{in}$. 
\item Obtain 
\begin{align}
\widehat{\mathbf{P}}=\frac{1}{2}\widetilde{\mathbf{P}}^\top_{in}(\widetilde{\mathbf{P}}_{11}^++\widetilde{\mathbf{P}}_{11}^{\top+})\widetilde{\mathbf{P}}_{in}, \label{ego:estimator}
\end{align}
where $\widetilde{\mathbf{P}}_{11}$ is the sub-matrix of $\widetilde{\mathbf{P}}_{in}$ that consists of the in-sample columns of $\widetilde{\mathbf{P}}_{in}$.
\end{enumerate}

\subsection{Interpretations of $\widehat{\mathbf{P}}$}
The low-rank approximation provides an interpretable parametrization in addition to an estimate of the probability matrix.   The rank $r$ approximation to  $\mathbf{P}$ can always be written as $\mathbf{P}_r=\mathbf{R}^\top\mathbf{Z}\mathbf{R}$, where $\mathbf{R}\in\mathbb{R}^{r\times N}$ and $\mathbf{Z}\in\mathbb{R}^{r\times r}$. 
Let $\mathbf{A}_{in}\stackrel{SVD}{=}\mathbf{U}\mathbf{D}\mathbf{V}^\top$ and $\widetilde{\mathbf{P}}_{in}\stackrel{SVD}{=}\mathbf{U}_r\mathbf{D}_r\mathbf{V}_r^\top$.
Correspondingly, we can rewrite \eqref{ego:estimator} as 
\begin{align*}
\widehat{\mathbf{P}}&=\mathbf{V}_r\mathbf{D}_r\mathbf{U}_r^\top\widehat{\mathbf{X}}\mathbf{U}_r\mathbf{D}_r\mathbf{V}_r^\top 
=\mathbf{V}_r(\mathbf{D}_r\mathbf{U}_r^\top\widehat{\mathbf{X}}\mathbf{U}_r\mathbf{D}_r)\mathbf{V}_r^\top 
:=\widehat{\mathbf{R}}^\top\widehat{\mathbf{Z}}\widehat{\mathbf{R}}.
\end{align*}
Thus, $\mathbf{V}_r$ gives an estimated embedding of the network in a space equipped with an inner product represented by the matrix $\widehat{\mathbf{Z}}$.

Similarly to the random dot product graph model \cite[e.g.][]{Sussman2012, Tang2014}, one can view the columns of $\mathbf{R}$ as coordinates of nodes in a pseudo-Euclidean space, with $p_{ij}$ determined by the inner product of points corresponding to nodes $i$ and $j$. The inner product is represented by $\mathbf{Z}$, and if desired, one can choose various additional contraints to impose on $\mathbf{Z}$ to fit a particular model.   Thus our estimator can be specialized to fit stochastic block models, dot-product models \citep{Young2007}, latent eigenmodels \citep{Hoff2007}, and hyperbolic models \citep{Krioukov2010, Albert2014}.

\section{Numerical evaluation}\label{sec:numerical_study}
\subsection{Tuning parameter selection}
We choose the approximation rank $r$, which can be viewed as a tuning parameter,via a general resampling scheme.   We repeatedly sample $k\in\{1,\dots,n\}$ and set $\mathbf{A}_{sub}$ be a submatrix of $\mathbf{A}_{in}$ by deleting the $k$-th row from $\mathbf{A}_{in}$. Although one can conduct leave-one-out cross-validation on rows, sampling $k$ rows at random reduces computational time when $n$ is large. 
Applying the proposed algorithm to $\mathbf{A}_{sub}$, we can estimate predictive accuracy by computing the area under the ROC curve (AUC) on the entries $\{A_{ki}: i\notin I\cup \{k\}\}$.   Alternatively, one could use a self-tuning or tuning-free method, such as universal singular value thresholding \citep{Chatterjee2012}
to obtain $\widehat{\mathbf{P}}_{in}$. This  will further reduce computational cost.

\subsection{Comparison with benchmarks}
Here we compare empirically our method  (SE, for subspace estimation) to several widely used algorithms for link prediction. We included the standard CUR decomposition (CUR) \citep{Mahoney2009} to show the importance of the subspace estimation step for egocentrically sampled networks.  
 From matrix completion methods with independently and identically sampled entries, we chose the universal singular value thresholding (USVT) and the nuclear norm regularization with inexact augmented Lagrange multiplier method (MC-IALM) \citep{Lin2010a}, two widely used and representative methods from their respective classes. For these two methods, we also show results on incomplete adjacency matrices with entries missing uniformly at random,  to show the effect of the egocentric sampling scheme on the performance of standard matrix completion methods. We also included the neighborhood smoothing method (NS) \citep{Zhang2015}, which is a graphon estimation method with demonstrated good performance on link prediction. The method uses a similarity measure between nodes, and given the incomplete nature of our network, we replaced $\mathbf{A}^2$ proposed by \cite{Zhang2015}  by $\mathbf{A}_{in}^\top\mathbf{A}_{in}$.    

\subsection{Synthetic networks}
First, we evaluate the performance of our method on simulated datasets. We generate networks from three common network models described in Table \ref{ego:table:generative_models}. For all networks, we set $N = 500$, first generate i.i.d.\ $X_i$'s for $i=1,\dots, N$ and then generate $A_{ij}\sim \mathrm{Bernoulli}(\phi f(X_i,X_j))$, where $\phi$ is a coefficient that controls the average degree. We tested our method and the benchmark methods under a range of sampling rates $\rho=n/500$ and average degree $d$.  In addition, we sampled pairs of nodes uniformly at random rather than whole rows, and applied the generic matrix completion methods USVT and MC to investigate how much the sampling scheme affects their performance. 
Table \ref{ego:table:generative_models} summarizes the simulation settings, and includes the rank of $\mathbf{P}$ and numerical rank of $\mathbf{A}$, defined as $\|\mathbf{A}\|_F^2/\|\mathbf{A}\|_2^2$, where $\|\cdot\|_F$ and $\|\cdot\|_2$ are the Frobenius norm and spectral norm, respectively.   Note that numerical rank of $\mathbf{A}$ increases with average degree.

\begin{figure}
\begin{subfigure}[b]{\linewidth}
\begin{adjustbox}{width=1\linewidth,center}
\includegraphics[scale=0.3]{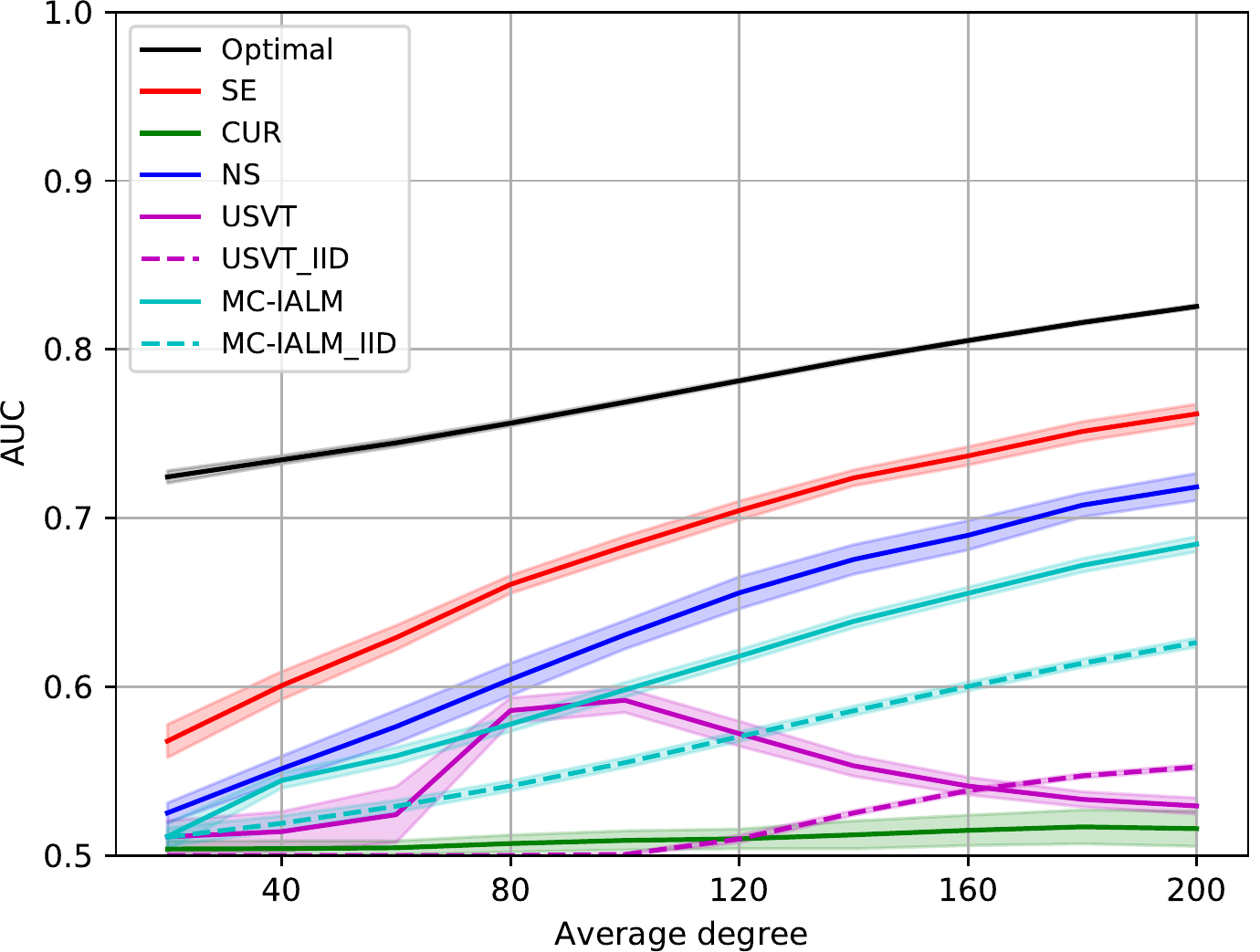}
\includegraphics[scale=0.3]{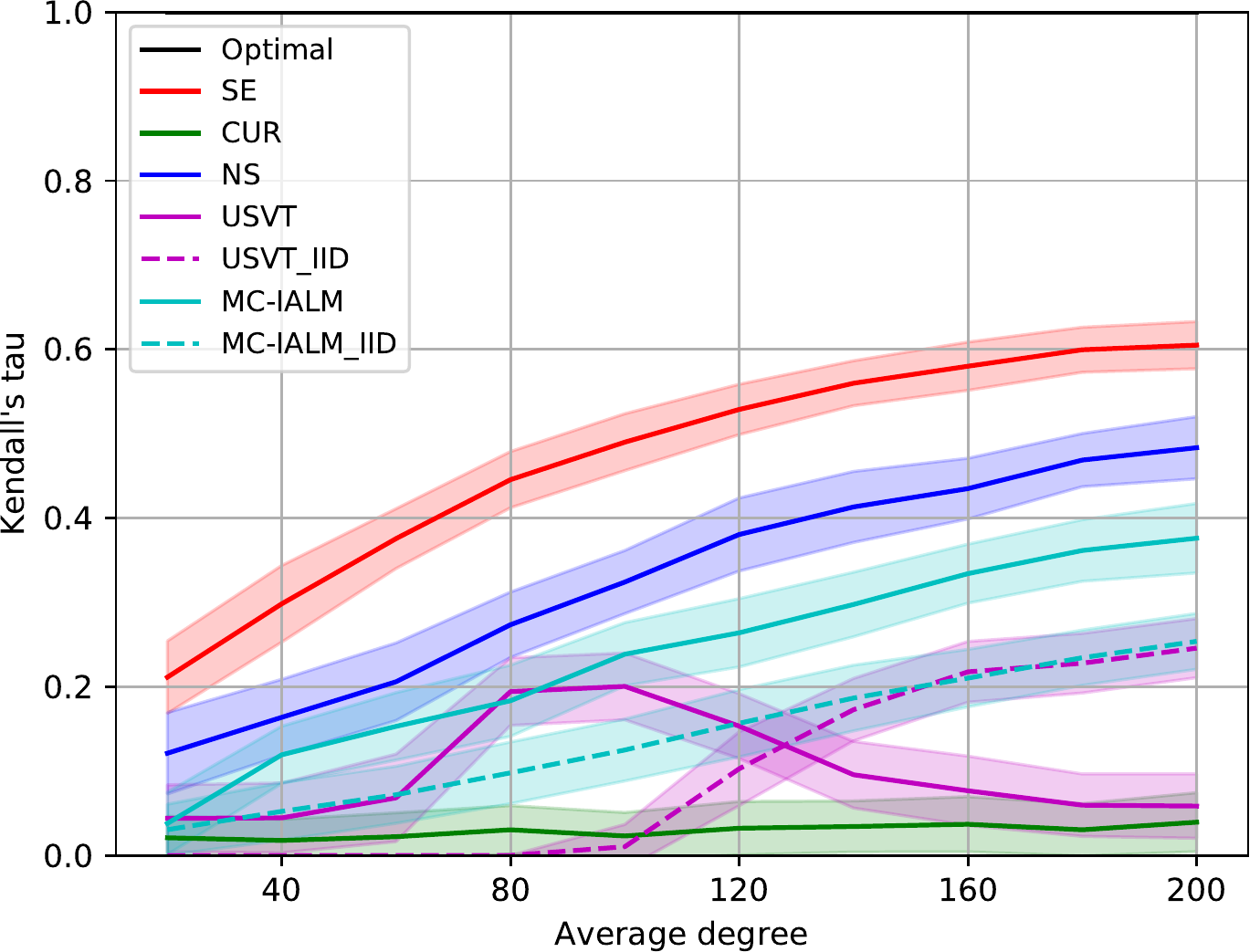} 
\end{adjustbox}
\caption{Distance model}\label{ego:fig:dist_deg}
\end{subfigure}

\bigskip
\begin{subfigure}[b]{\linewidth}
\begin{adjustbox}{width=1\linewidth,center}
\includegraphics[scale=0.3]{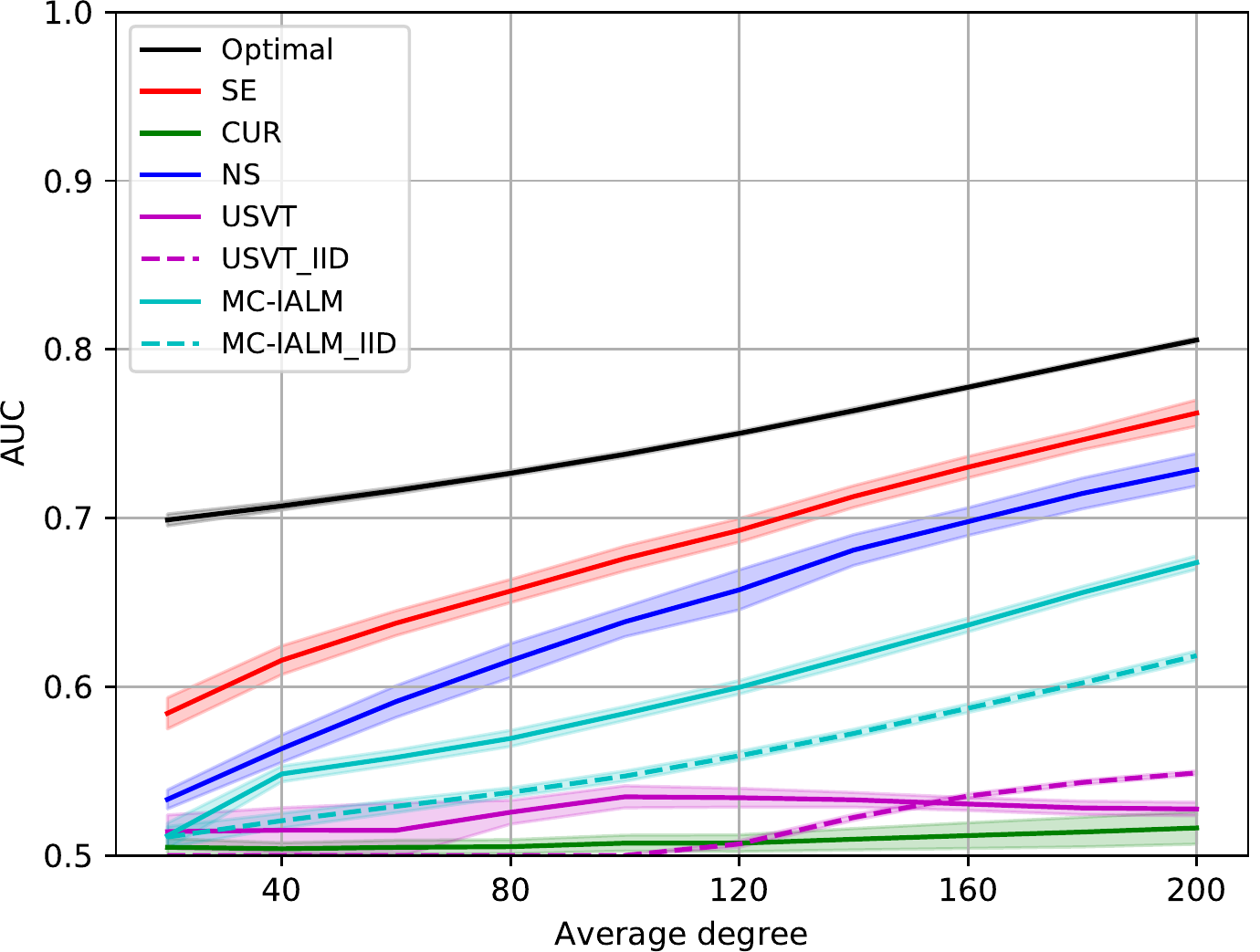}
\includegraphics[scale=0.3]{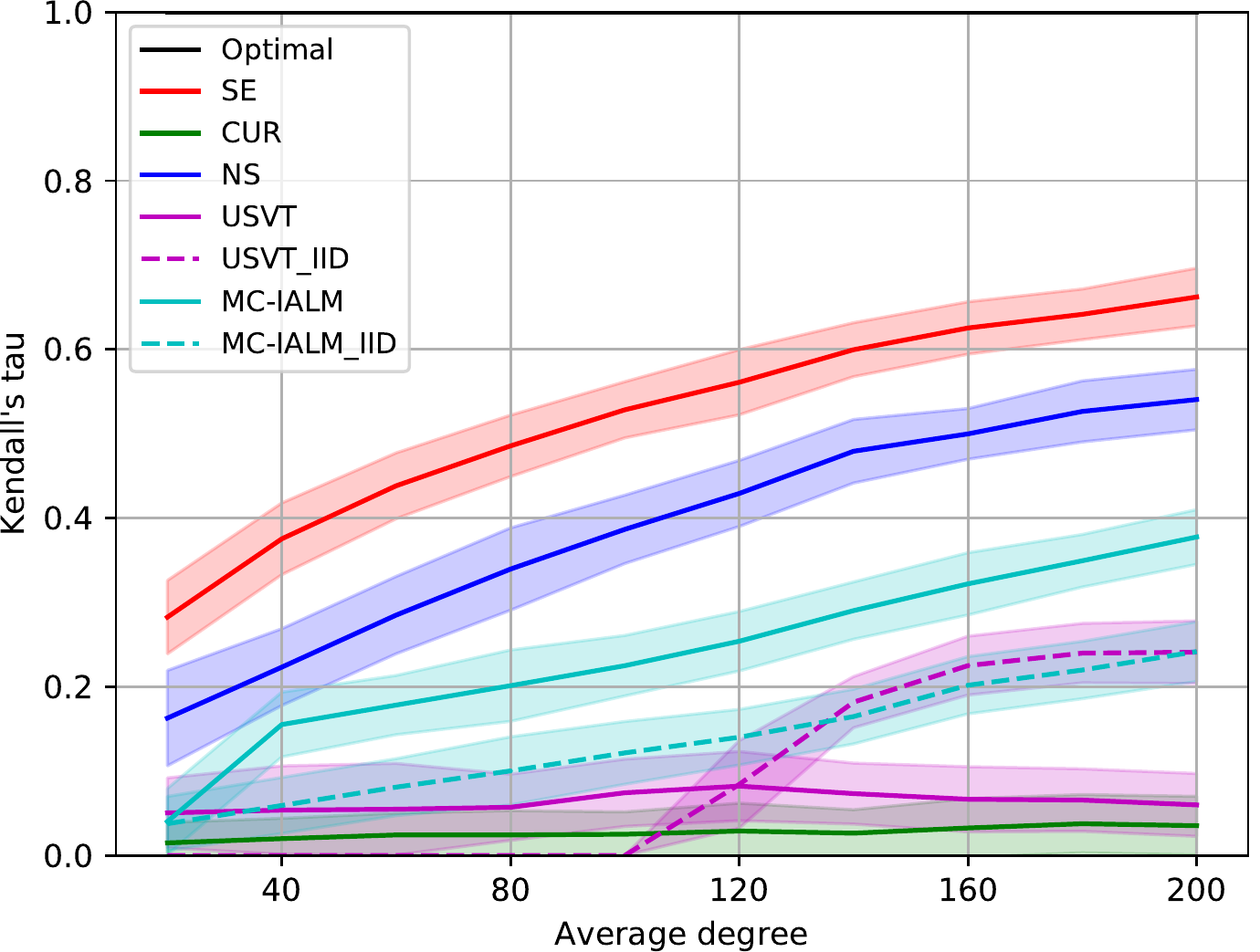} 
\end{adjustbox}
\caption{Product models}\label{ego:fig:prod_deg}
\end{subfigure}

\bigskip
\begin{subfigure}[b]{\linewidth}
\begin{adjustbox}{width=1\linewidth,center}
\includegraphics[scale=0.3]{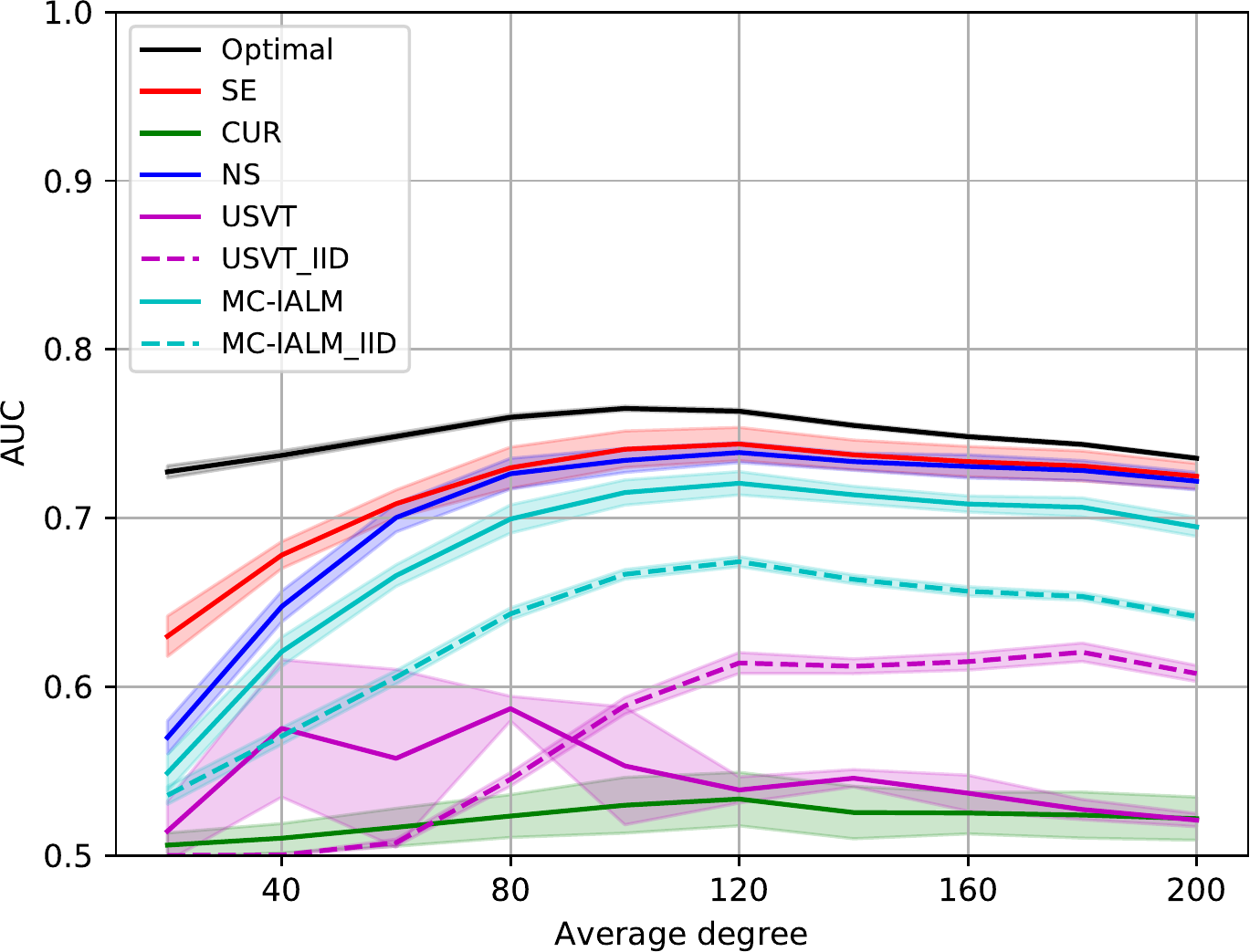}
\includegraphics[scale=0.3]{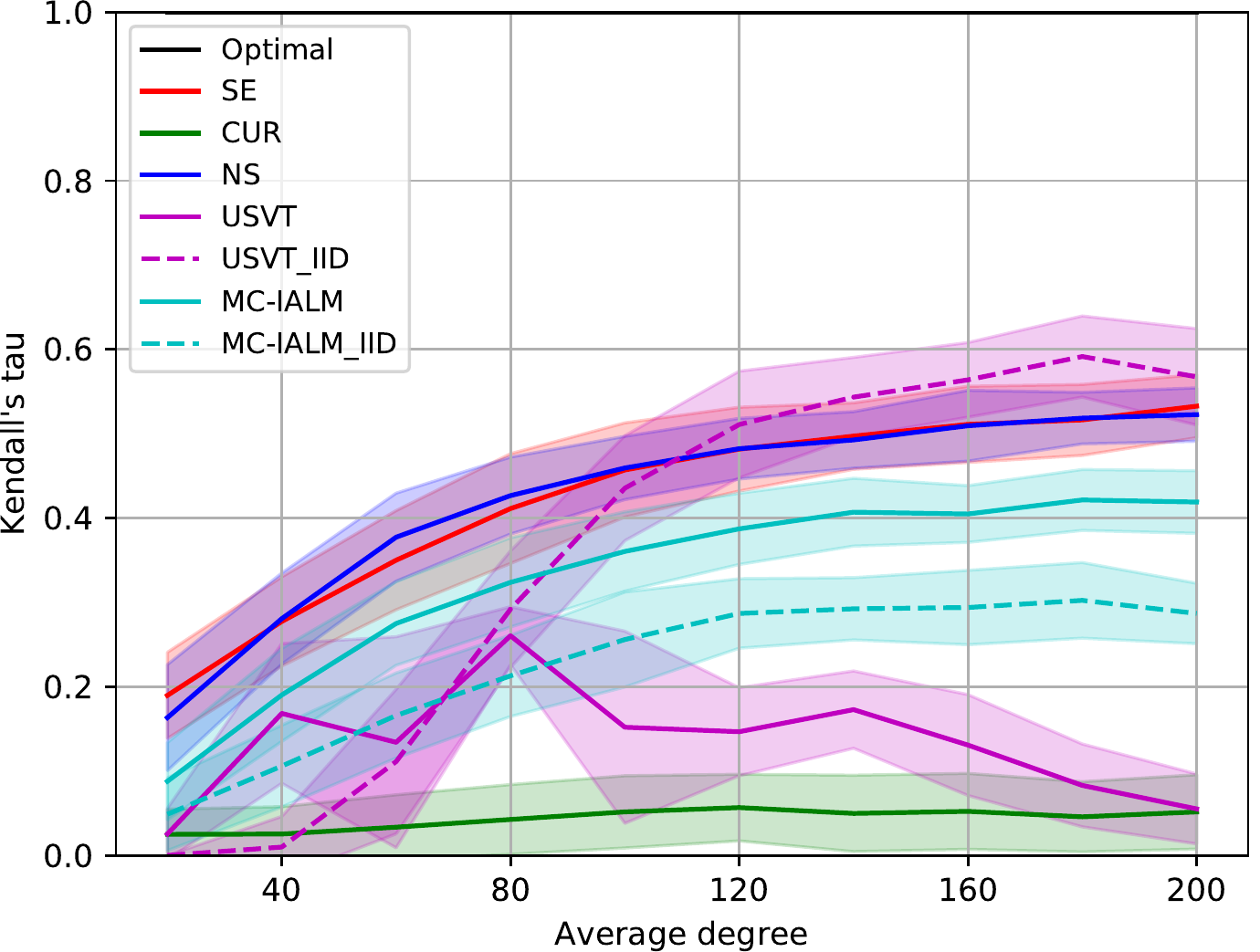} 
\end{adjustbox}
\caption{SBM}\label{ego:fig:sbm_deg}
\end{subfigure}
\caption{Predictive AUC and Kendall's tau as average degree varies and sampling rate $\rho = 0.2$, with confidence bands of $\pm 1$ SE. }
\end{figure}

\begin{figure}
\begin{subfigure}[b]{\linewidth}
\begin{adjustbox}{width=1\linewidth,center}
\includegraphics[scale=0.3]{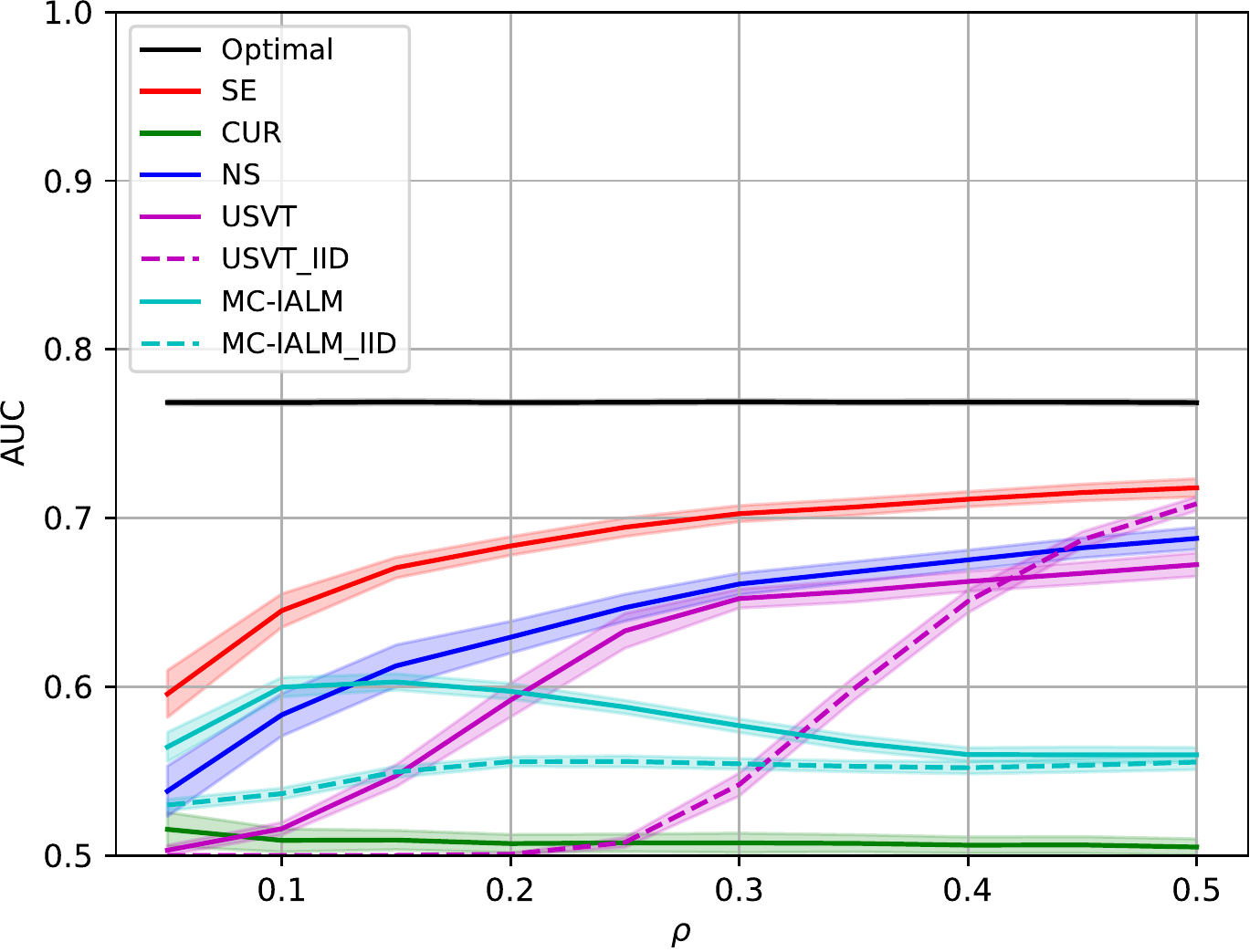}
\includegraphics[scale=0.3]{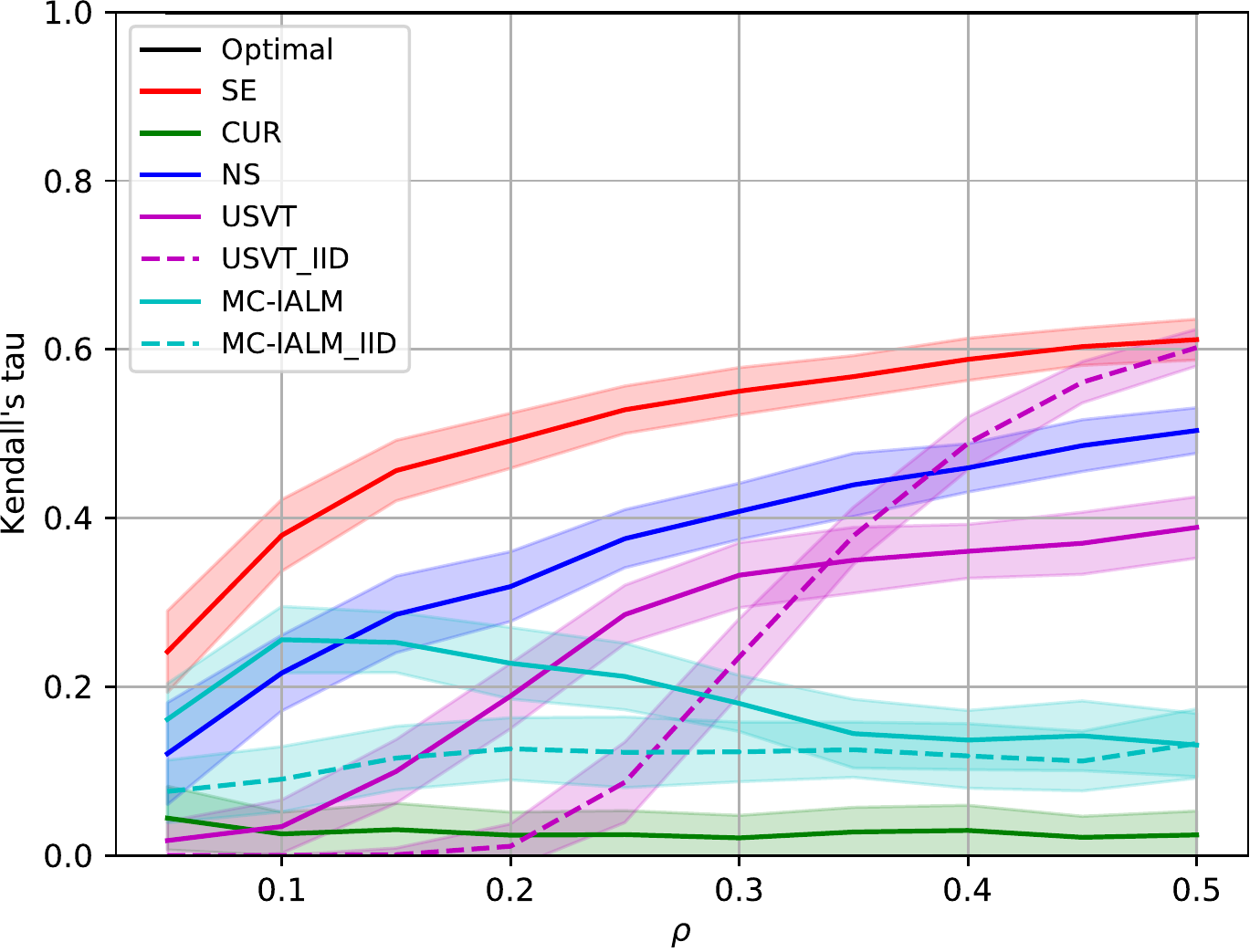} 
\end{adjustbox}
\caption{Distance model}\label{ego:fig:dist_rho}
\end{subfigure}

\bigskip
\begin{subfigure}[b]{\linewidth}
\begin{adjustbox}{width=1\linewidth,center}
\includegraphics[scale=0.3]{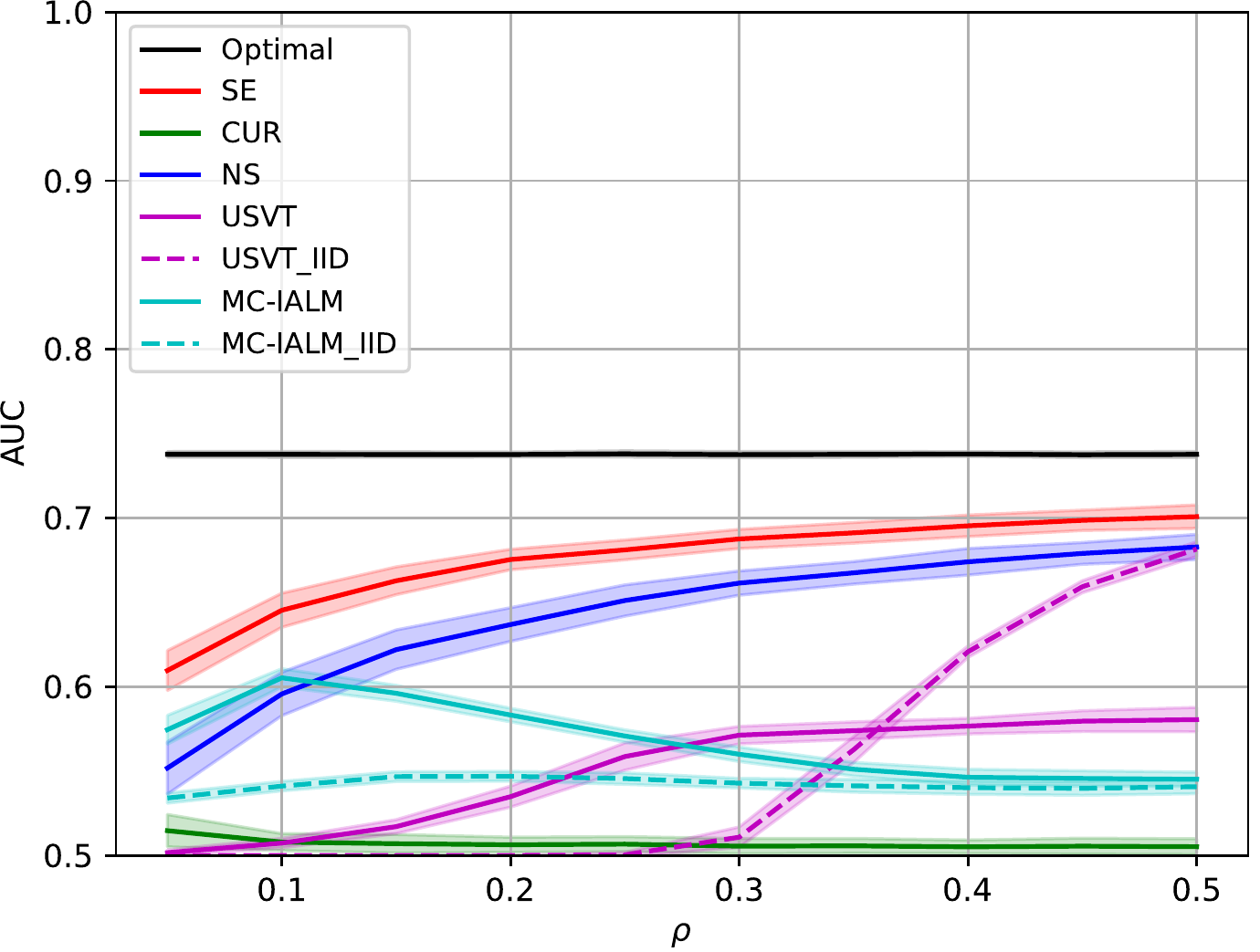}
\includegraphics[scale=0.3]{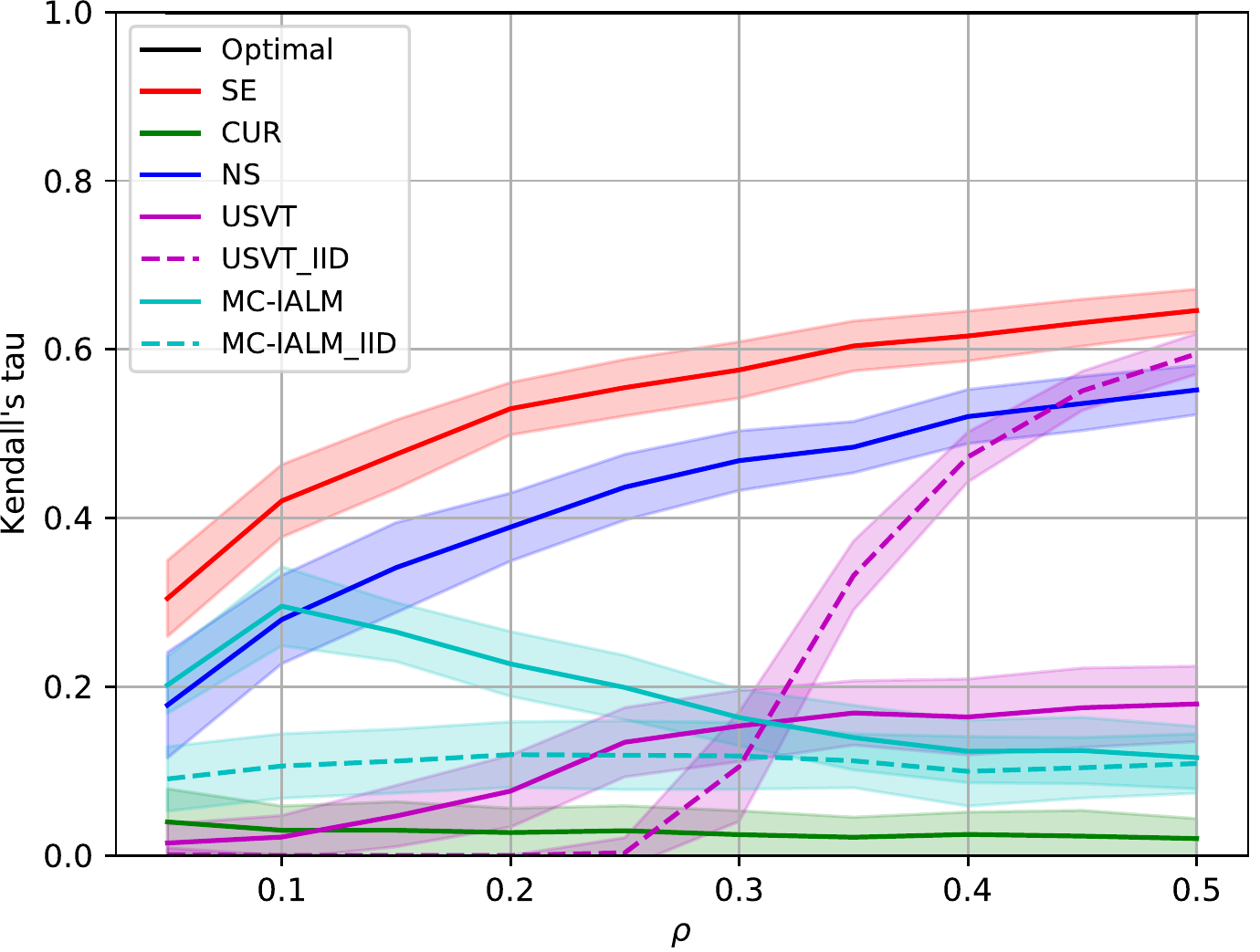} 
\end{adjustbox}
\caption{Product models}\label{ego:fig:prod_rho}
\end{subfigure}

\bigskip
\begin{subfigure}[b]{\linewidth}
\begin{adjustbox}{width=1\linewidth,center}
\includegraphics[scale=0.3]{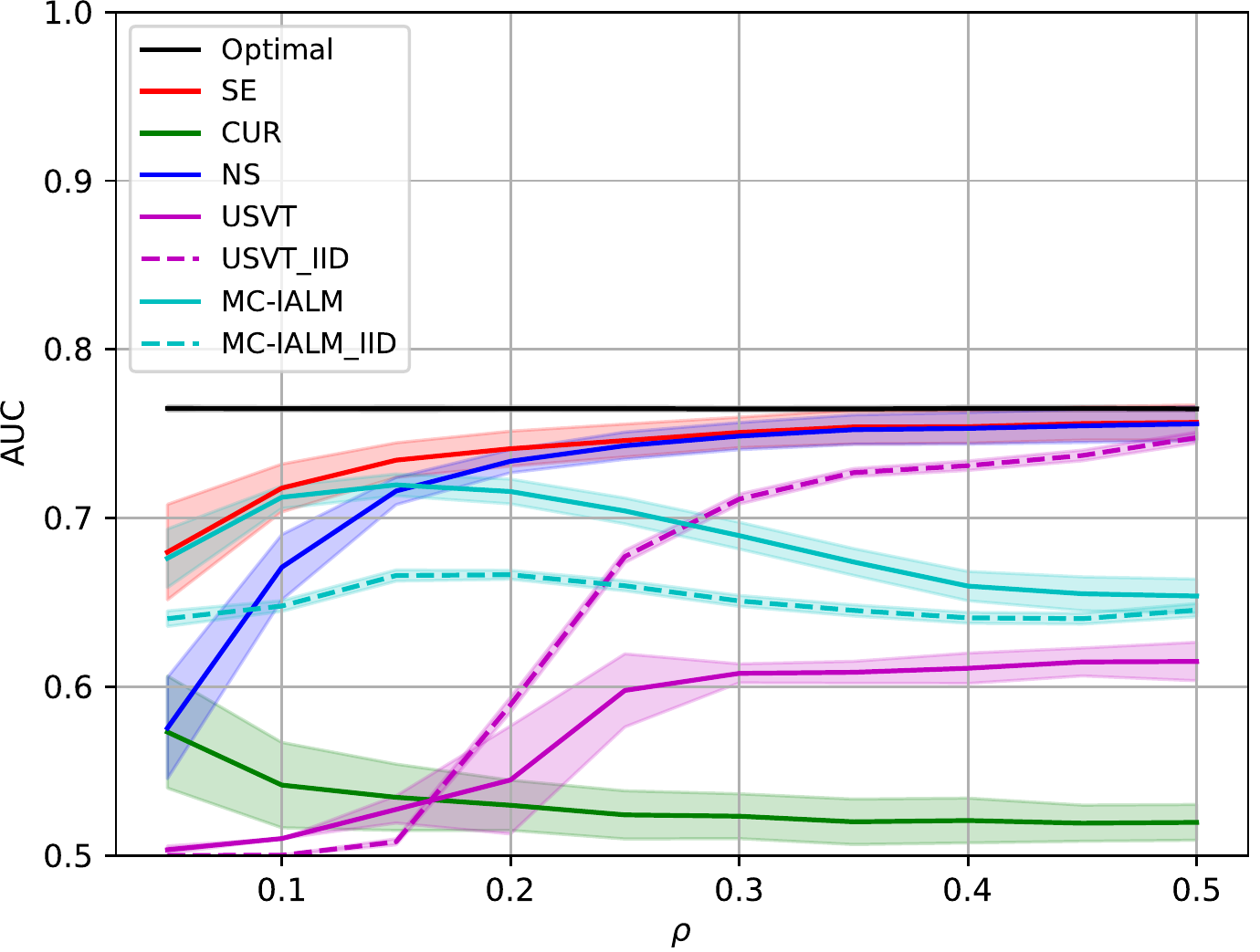}
\includegraphics[scale=0.3]{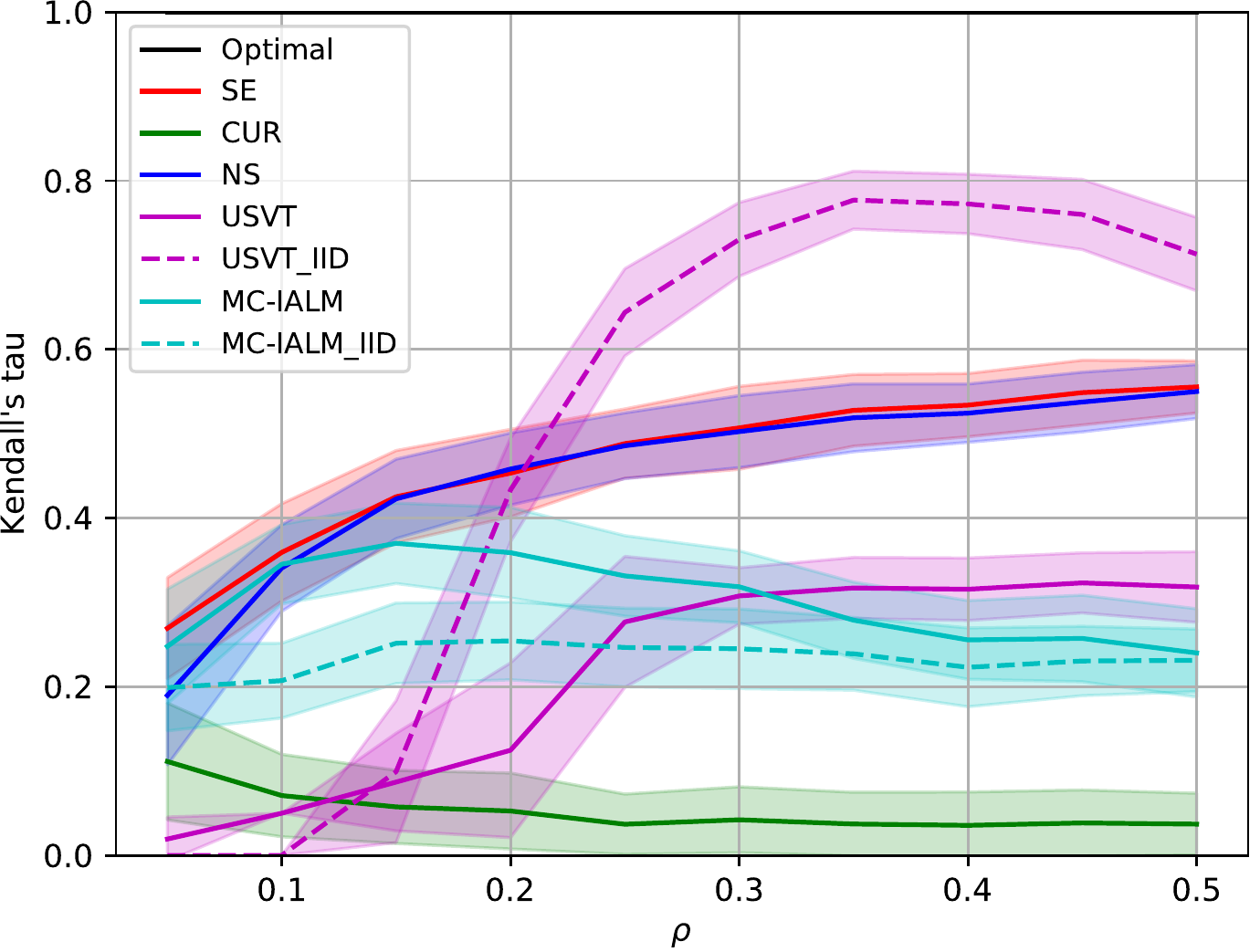} 
\end{adjustbox}
\caption{SBM}\label{ego:fig:sbm_rho}
\end{subfigure}
\caption{Predictive AUC and Kendall's tau with as the sampling rate $\rho$ varies and average degree $d = 100$, with confidence bands of $\pm 1$ SE.   }
\end{figure}

We vary the settings in  two ways:   varying the sampling fraction $\rho$ from $0.05$ to $0.5$ while keeping the average degree fixed at $d = 100$ (Figures \ref{ego:fig:dist_deg}-\ref{ego:fig:sbm_deg}), and varying the average degree $d$ from 10 to 200 while keeping the sampling fraction fixed at $\rho = 0.2$ (Figures \ref{ego:fig:dist_rho}-\ref{ego:fig:sbm_rho}). 
We also measure link prediction performance  two different ways:  by predictive area under the ROC curve (AUC), defined as 
\[\mathrm{AUC}(\widehat{\mathbf{P}},\mathbf{A})=\frac{\sum_{i,j,i',j'\notin I}1(A_{ij}=1,A_{i'j'}=0, \widehat{p}_{ij}>\widehat{p}_{i'j'})}{\sum_{i,j,i',j'\notin I}1(A_{ij}=1,A_{i'j'}=0)} \ , \]
and by predictive Kendall's tau, defined as 
\[\tau(\widehat{\mathbf{P}},\mathbf{P})=\frac{2\sum_{i,j,i',j'\notin I}1(p_{ij}>p_{i'j'}, \widehat{p}_{ij}>\widehat{p}_{i'j'})}{\sum_{i,j,i',j'\notin I}1(p_{ij}>p_{i'j'})}-1 . \]
We compute both measures on the unobserved sub-matrix $[A_{ij}]_{i,j\in\mathcal{I}}$. The results for each setting are  averaged over 100 replications.

\begin{table}
\caption{Generative models for synthetic networks}\label{ego:table:generative_models}
\begin{adjustbox}{center}
\begin{tabular}{l|lcccc}
%\toprule
 Model & Dist of $X_i$ & $p_{ij}\propto f(X_i,X_j)$ & Rank($\mathbf{P}$) & Ave. deg.& Num. rank of $ \mathbf{A}$ \\
\midrule
Distance & $N_5(0, 1)$ & $(1 + e^{(\|X_i-X_j\|)})^{-1}$ & Full  &  33.9 & 1.9--14.7\\
Product & $\beta_5(0.5, 1)$ & $X_i^\top X_j$ & 5 & 52.8 & 1.7--14.2 \\
SBM & U$(1,\dots,5)$ & $.05+\frac{i-0.3}{6}1(i=j)$ & 5 & 60.7 & 2.6--12.6  \\
%\bottomrule
\end{tabular}
\end{adjustbox}
\end{table}

As the average degree grows (Figures \ref{ego:fig:dist_deg}-\ref{ego:fig:sbm_deg}),  our method, CUR, and NS all improve. Our method uniformly outperforms all benchmarks in terms of both predictive AUC and Kendall's tau for distance and product models.   For the SBM, the graphon method NS also gives comparable accuracy (except in the AUC measure for small $d$), possibly because the block structure allows NS to find sufficiently many similar neighborhood nodes.  
Figures \ref{ego:fig:dist_rho}-\ref{ego:fig:sbm_rho} show that increasing the sampling rate improves performance of all methods, as one would expect. Our method again outperforms all benchmarks on egocentrically sampled networks.   NS performs similarly on SBM again, except it performs somewhat worse for small values of $\rho$.     Additionally, comparing the i.i.d.\ versions of matrix completion and USVT to their results on egocentrically sampled networks, we find that which scenario gives better results depends on the specific values of $\rho$ and $d$ for USVT, whereas ghd MC actually performs worse under the i.i.d.\ scenario.   Either way, they are almost always quite different and thus any results established under the i.i.d.\ scenario cannot be expected to carry over.   

\subsection{Real networks}

\begin{figure}
\begin{center}
\includegraphics[scale=0.57]{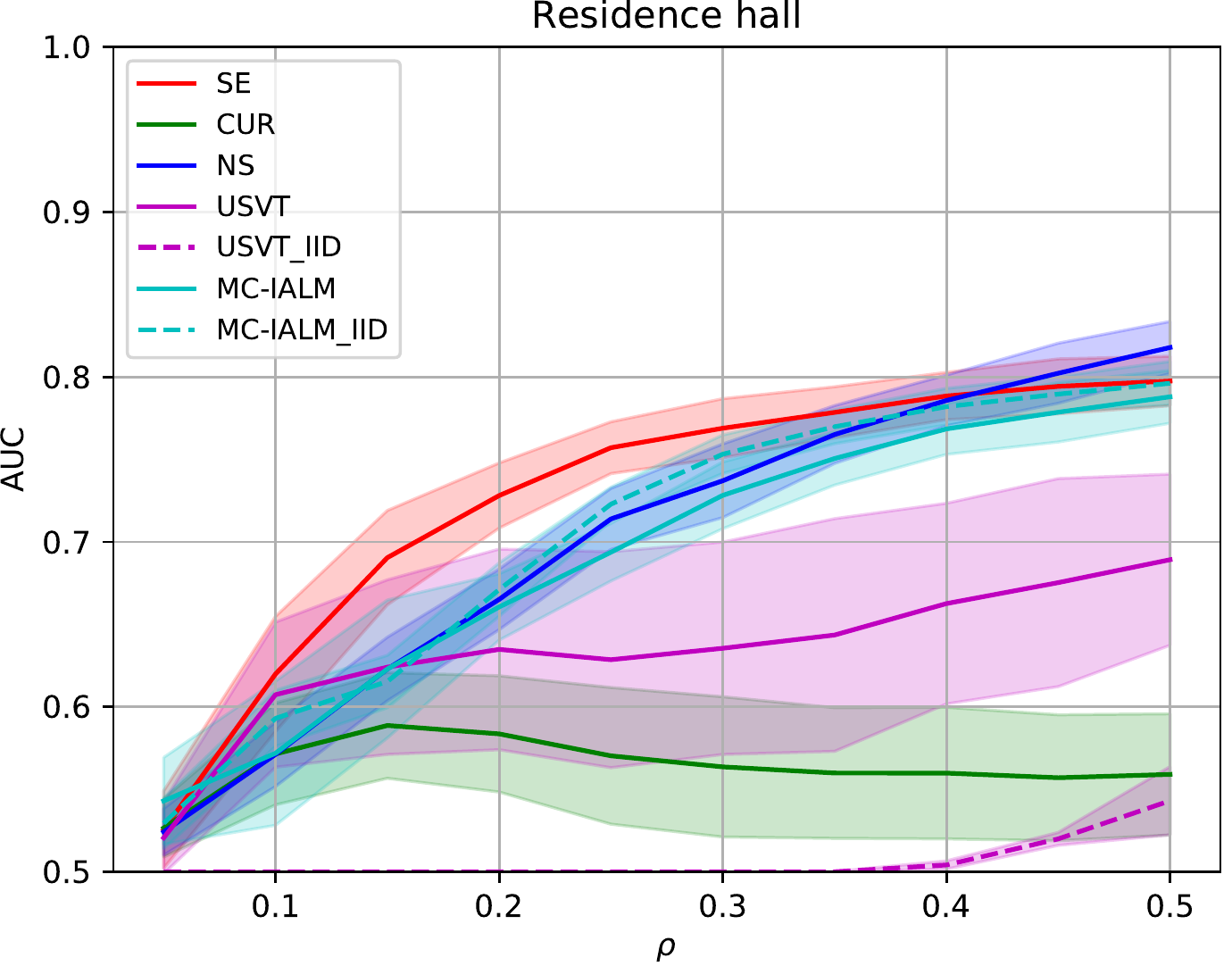}
\includegraphics[scale=0.57]{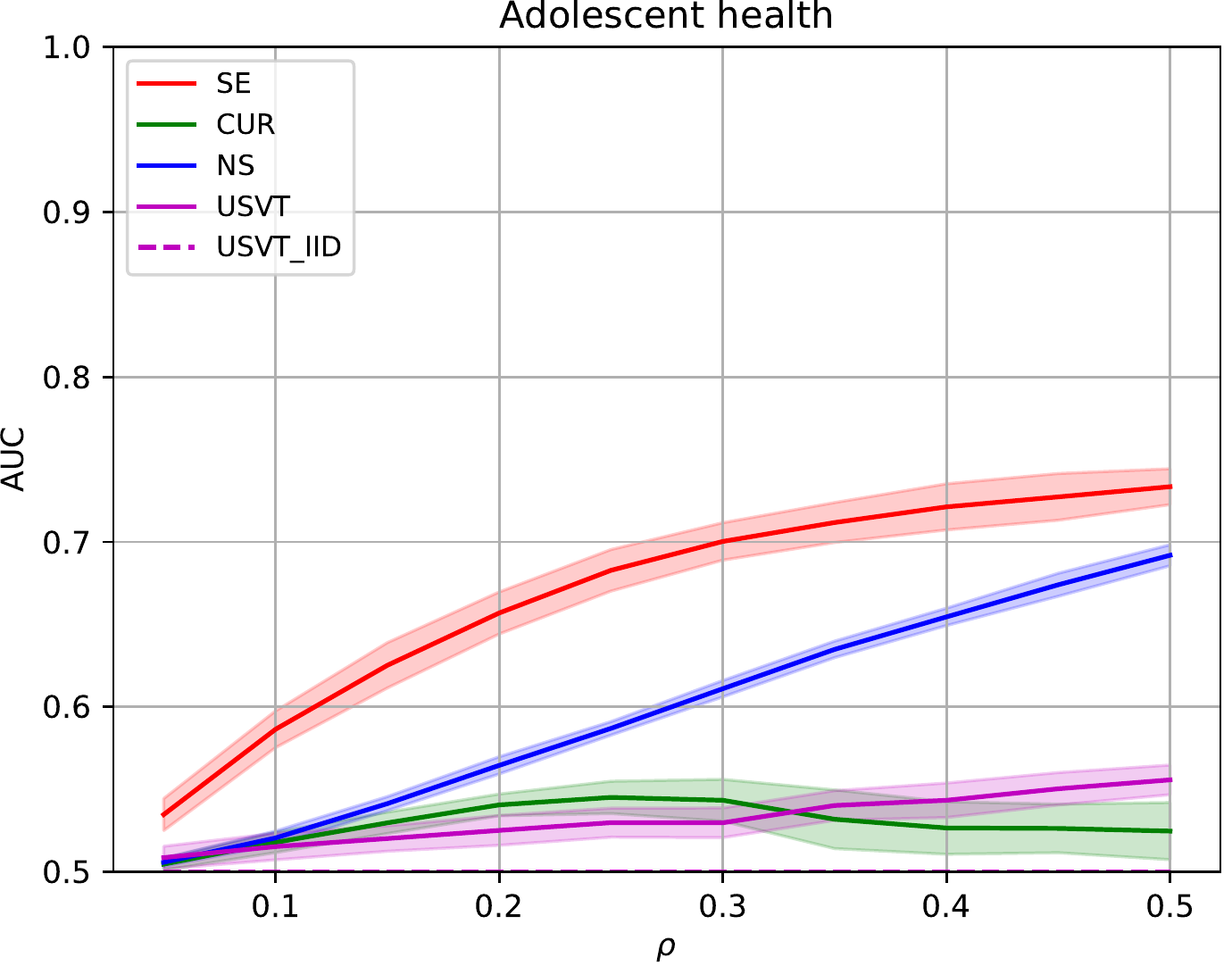} 
\includegraphics[scale=0.57]{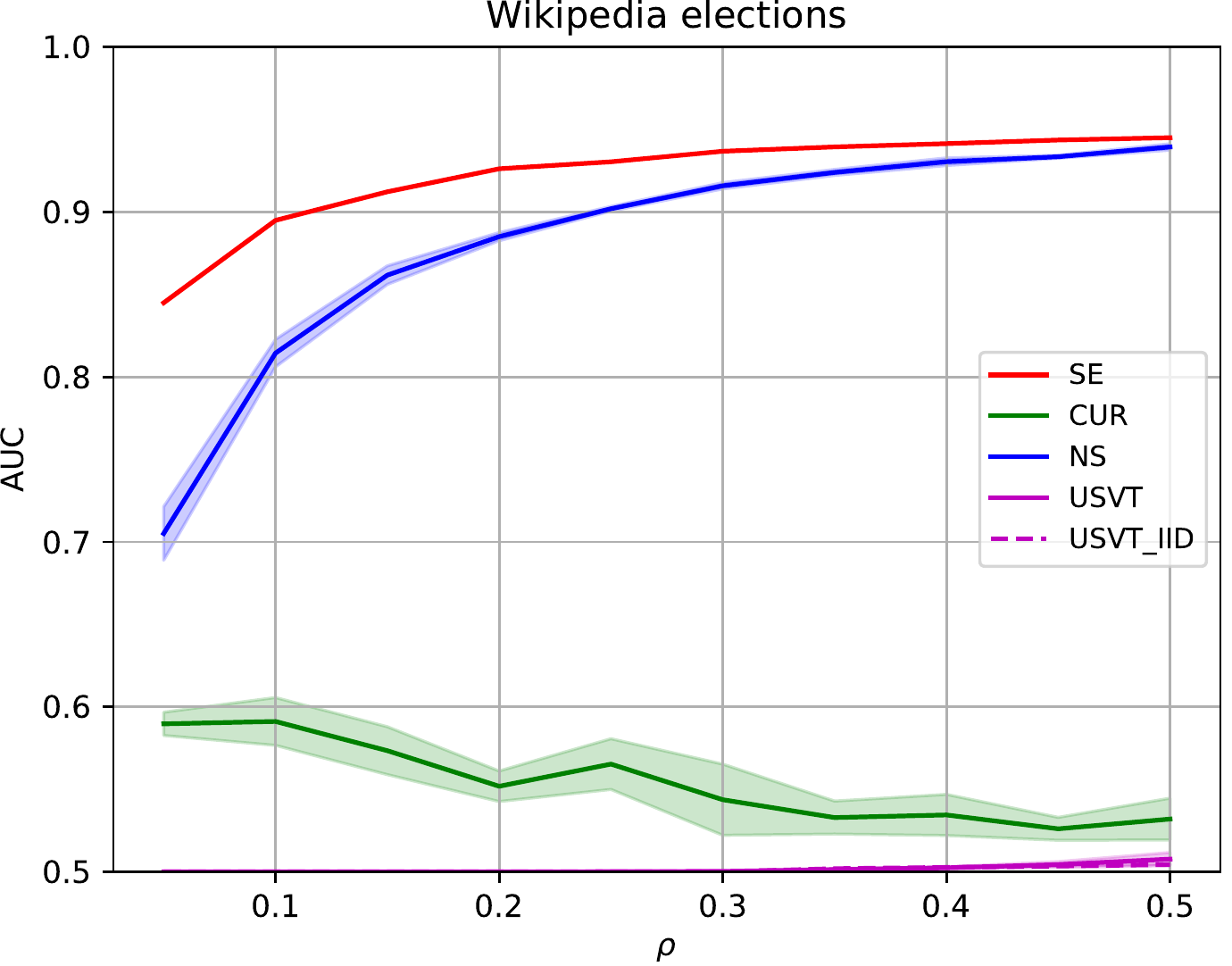}
\end{center}
\caption{Predictive AUC for real datasets with confidence bands of $\pm 1$ SE. }
\label{ego:fig:data_deg}
\end{figure}

Finally, we applied our method and the benchmark methods to the residence hall network \citep{Freeman1998}, which contains friendship between residents at a residence hall, the adolescent health network \citep{Moody2001}, which contains a social network between students in a survey, and the Wikipedia election network \citep{Leskovec2010a}, in which dataset nodes represent users and links represent voting for each other in admin elections. 
  Summaries of these three social networks are given in Table \ref{ego:table:datasets}. We sampled 5\% to 50\% of nodes to construct egocentric samples and computed predictive AUC on unobserved pairs of nodes. Since the true $\mathbf{P}$ is not available for data, we did not compute Kendall's tau.  Note that some benchmark methods had to be omitted for some datasets due to their high computational cost.   The results in Figure \ref{ego:fig:data_deg} closely agree with what we found in simulations, with a simliar trend as a function of $\rho$ and our method outperforming the benchmarks, particularly for small $\rho$.

\begin{table}
\caption{Descriptive statistics of datasets}\label{ego:table:datasets}
\begin{center}
\begin{tabular}{l|cccc}
\toprule
  Dataset & $N$ & $m$ & Avg. deg. & Num. rank \\
\midrule
Residence hall & 217 & 2672 & $24.6$ & 7.91 \\
Adolescent health &  2539 & 12969 & $10.2$ & 119.44  \\
Wikipedia elections & 7118 & 103675 & $28.3$ & 10.57 \\
\bottomrule
\end{tabular}
\end{center}
\end{table}

\section{Discussion}\label{sec:discussion}

To sum up, our method achieved a better accuracy for link prediction in egocentrically sampled networks than other benchmarks, and its computational cost is only slightly more than that of the CUR decomposition.   Our method appears to have the biggest advantage in the hardest scenarios of small sampling rates or sparse networks, making it useful in practice for survey-based data often collected in various social studies.   

The method works by exploiting a presumed low-rank structure of the underlying probability matrix and employing subspace estimation.   While that general idea also underlies many matrix completion methods, our method performs much better than generic matrix completion, showing that the egocentrically sampled networks do require a careful and separate treatment.   Further, since subspace estimation denoises the observed matrix, essentially performing matrix completion on the sampled rows as a first step, we can expect our method to handle missing values in the sampled rows relatively well.

Promising directions for future work include design of survey procedures with importance sampling, if there is prelimiary data from which one can pre-compute node weights (this is done in the CUR algorithm, since it is aimed at matrix compression and assumes the entire matrix is initially available, which would not be the case in a survey).   Another future direction is to investigate link prediction under other sampling schemes, such as snowball sampling.

\renewcommand*{\bibname}{References}
\bibliographystyle{apalike}
\bibliography{ref}

\end{document}